\documentclass[a4paper,12pt,3p]{elsarticle}

\usepackage[english]{babel}
\usepackage{amssymb}
\usepackage{color}

\newcommand{\rmd}{\mathrm{d}}
\newcommand{\eqref}[1]{(\ref{#1})}

\makeatletter
\def\ps@pprintTitle{%
 \let\@oddhead\@empty
 \let\@evenhead\@empty
 \def\@oddfoot{}%
 \let\@evenfoot\@oddfoot}
\makeatother

\begin{document}

\title{Stochastic Model of Financial Markets Reproducing Scaling and Memory in Volatility Return Intervals}

\author[boston,itpa]{V. Gontis\corref{vg}}
\ead{vygintas@gontis.eu}

\author[boston,isr]{S. Havlin}

\author[itpa]{A. Kononovicius}

\author[boston,rijeka,zagreb]{B. Podobnik}

\author[boston]{H. E. Stanley}

\cortext[vg]{Corresponding author}

\address[boston]{Center for Polymer Studies and Department of Physics, Boston University, Boston, MA 02215, US}
\address[itpa]{Institute of Theoretical Physics and Astronomy, Vilnius University, Vilnius, LT 10222, Lithuania}
\address[isr]{Department of Physics, Bar-Ilan University, Ramat Gan, IL 52900, Israel}
\address[rijeka]{Faculty of Civil Engineering, University of Rijeka, Rijeka, HR 51000, Croatia}
\address[zagreb]{Zagreb School of Economics and Management, Zagreb, HR 10000, Croatia}

\begin{abstract}
We investigate the volatility return intervals in the NYSE and FOREX
markets. We explain previous empirical findings using a model based
on the interacting agent hypothesis instead of the widely-used efficient
market hypothesis. We derive macroscopic equations based on the microscopic
herding interactions of agents and find that they are able to reproduce various stylized
facts of different markets and different assets with the same set of model parameters. We show that the
power-law properties and the scaling of return intervals and other
financial variables have a similar origin and could be a result of a general
class of non-linear stochastic differential equations derived from a
master equation of an agent system that is coupled by herding
interactions. Specifically, we find that this approach enables us to recover the volatility return interval statistics as well as volatility probability and spectral  densities  for the NYSE and FOREX markets, for different assets, and for different time-scales. We find also that the historical S\&P500 monthly series exhibits the same volatility return interval properties recovered by our proposed model. Our statistical results suggest that human
herding is so strong that it persists even when other evolving
fluctuations perturbate the financial system.
\end{abstract}


\maketitle

\section{Introduction}
To estimate risk in a financial market it is essential that we
understand the complex market dynamics involved
\cite{Bouchaud2004Cambridge,Sornette2004Princeton}.  Statistical physics 
has been found useful dealing with the general concepts of complexity 
and its applications in finance \cite{Karsai2012NIH,Chakraborti2011RQUF1,Gabaix2009AR}. Financial markets are among the most interesting examples of such complex social systems where methods of statistical physics face extreme challenges \cite{Farmer2012EPJ}. Although our current understanding of financial fluctuations and the nature of
microscopic market interactions remains limited and ambiguous
\cite{Shiller2014AER,Kirman2014MD}, as vast amounts of financial data
have become more available we are now able to apply advanced methods of
empirical analysis to gain greater insight into the market's complexity
\cite{Campbell1996Princeton,Mantegna2000Cambridge,Bouchaud2004Cambridge,Sornette2004Princeton}.

Here we use a general agent-based stochastic model
\cite{Gontis2014PlosOne}, reproducing first and second order statistics of absolute return in the financial markets and find that with very minor modifications it is able  to reproduce various statistical properties of the high volatility return intervals
\cite{Yamasaki2005PNAS,Wang2006PhysRevE,Wang2008PhysRevE,Bunde2011EPL,Bunde2014PRE}.

We focus on the heuristic model of volatility, which is defined as fluctuations in the absolute
returns, across a wide range of time-scales from one minute to one month. There are many other attempts of econometric approach to the problem of behavioral opinion dynamics of agents in the financial markets \cite{Brock2001RES,Diks2005JEDC,Franke2012JEDC,Lux2012JEDC,Goldbaum2014JEBO,He2015JEF,Jang2015CE} able to explain fat tails and volatility clustering. Usually these econometric analyses based on generalized or simulated
method of moments (GMM or SMM) are limited to the oversimplified agent models with small number of parameters. To our knowledge, the values of parameters in these models are dependent on selected time window of return definition and are not universal for other time scales. Earlier proposed model of the financial markets \cite{Gontis2014PlosOne}, which we use here, accumulates some general features of agent dynamics and price formation from Ref. \cite{Alfarano2005CompEco,Alfarano2008Dyncon}. This model further generalizes herding dynamics for the three groups of agents \cite{Kononovicius2013EPL} by the continuous stochastic differential equations derived for the infinite number of agents with pairwise global interactions. At the same time the proposed model is able to account for the
feedback of market volatility on the market trading activity observed in the financial markets
\cite{Gabaix2003Nature,Farmer2004QF,Gabaix2006QJE,Gontis2006JStatMech,Gontis2007PhysA,Podobnik2009PNAS,Scheinkman2014CUP}. The main task of this work is to  demonstrate that proposed stochastic model with the same set of parameters allows
to understand statistics of absolute return intervals for wide range of time and threshold scales even when the values are extreme.  

We find that the statistical properties of return intervals
are universal for a broad range of financial markets,
from NYSE and FOREX.  The model can reproduce these statistical
properties by using the same set of parameters for varying time-scales,
from high frequency data to monthly S\&P500 index values across a
145-year period \cite{Shiller2015IE}.  These results imply that the
various power-law statistics of financial markets might be due to a
non-linear stochasticity, which we incorporate into the herding-based
model of financial markets \cite{Ruseckas2014JSM,Kononovicius2015PhysA}. Though the proposed model is designed to analyze statistical properties of volatility and the price of assets is not considered, the revealed bursting behavior  extends our understanding of bubbles in financial markets  \cite{Shiller2015IE,Scheinkman2014CUP} in general.

\section{Method}
We use a modified version of the three-state agent-based model
\cite{Gontis2014PlosOne,Kononovicius2013EPL} to reproduce and explain
the origin of the statistical properties of volatility return intervals
\cite{Yamasaki2005PNAS,Wang2006PhysRevE,Wang2008PhysRevE}. The interplay
between the endogenous dynamics of agents and exogenous noise is the
primary mechanism responsible for the observed statistical
properties. By exogenous noise we mean order
flow fluctuations.

Though our approach to the financial markets \cite{Gontis2014PlosOne,Kononovicius2013EPL} inherits  some essential features from herding based modeling proposed in \cite{Alfarano2005CompEco,Alfarano2008Dyncon} and other numerous papers, there are few significant extensions and different model interpretations we use in our approach. Let us shortly summarize our main assumptions: 
\begin{enumerate}
\item Pairwise global herding interactions of agents (traders) are assumed as the result of the pairwise interactions of traders during their trade actions. This conditions macroscopic description of agents by SDEs independent from the total number of agents, and macroscopic state feedback on the microscopic trading activity of agents.
\item The clustering of volatility and trading activity, long-range dependence and multifractality are related with the nonlinear nature of SDEs derived for corresponding financial variables.
\item The model has to incorporate endogenous (agent based) and exogenous (order flow) fluctuations as they coexist and interplay in the real markets.
\item There are at least three different time scales of return fluctuations in the financial markets a) the long term fluctuations of fundamentalists and chartists; b) the short term fluctuations of optimists and pessimists; c) the most frequent fluctuations of return related with order flow. 
\end{enumerate}
These assumptions lead to the consentaneous microscopic and macroscopic model combining endogenous agent based dynamics with stochastic dynamics driven by exogenous noise. We use visual empirical test here based on a double logarithmic axes histograms to select 9 independent model parameters seeking to reproduce many different power-law statistical properties at the same time. The heuristic consideration of noises generated by derived SDEs makes this parameter selection procedure preferable against formal fitting methods and helps to reproduce many stylized facts based on first and second order statistics with the same set of parameters for different markets and for different time windows of return  definition. 
\subsection{Endogenous versus exogenous}
The standard price model \cite{Jeanblanc2009Springer} and autoregressive conditional
heteroskedasticity (ARCH) family of
models \cite{Engle1982Econometrica,Bollerslev1986Econometrics} serve as
phenomenological frameworks consistent with endogenous volatility and
exogenous noise. For example, by analogy with ARCH family models we can
assume that the log return $r_{\delta}(t)=\ln P(t)-\ln P(t-\delta)$ of
the market price $P(t)$, defined at any moment $t$ for a time interval
$\delta$ can be modeled as a product of endogenous volatility
$\sigma(t)$ and exogenous noise $\omega(t)$
\begin{equation}
r_{\delta}(t) = \sigma(t) \omega(t).
\label{eq:return}
\end{equation}
Here for the sake of simplicity we use a Gaussian noise $\omega(t)$, and
volatility $\sigma(t)$ is assumed to be a linear function of the
absolute endogenous log price $\vert p(t) \vert = \vert \ln \frac{P(t)}{P_f} \vert$
\begin{equation}
\sigma(t) = b_0(1+ a_0 \vert p(t) \vert),
\label{eq:defvolatil}
\end{equation}
where $p(t)$ can be derived from the agent-based model (ABM) defining the ratio of market price $P(t)$ to fundamental price $P_f$
\cite{Gontis2014PlosOne}.  Here $b_0$ serves as a normalization
parameter, while $a_0$ determines the impact of endogenous dynamics on
the observed time series. Our model, defined by Eqs.~(\ref{eq:return})
and (\ref{eq:defvolatil}), comprises both the dynamic part described by
$\sigma(t)$ and the purely stochastic part described by $\omega(t)$. 

The motion of the financial Brownian particle colliding with the flow of limit orders in the real financial market \cite{Takayasu2015PRE} probably serves as a possible physical interpretation of the Gaussian noise in Eq. (\ref{eq:return}). The selected time window $\delta$ here is limited by the requirement that the change of $\sigma_t$ has to be inconsiderable. This means that exogenous fluctuations in this model are much more frequent than endogenous. Note that Eq. (\ref{eq:return}) in econometric consideration does not include any limits for $\delta$ as $\sigma(t)$ there is not related to the similar physical interpretations and is just formally defined through the auto-regressive model.

\subsection{ABM}
We use a version of the three-state agent-based herding model
\cite{Gontis2014PlosOne,Kononovicius2013EPL} to describe the endogenous
dynamics of agents in the financial markets and to reproduce the
statistical properties of volatility return intervals
\cite{Yamasaki2005PNAS,Wang2006PhysRevE,Wang2008PhysRevE}.

Agents interact globally as the pairwise interactions of traders during their trade actions are assumed. This assumption helps to overcome the problem of spacial structure of interactions usually considered in agent modeling approaches \cite{Ausloos2015Springer} and allows to account for the observed relation of return with trading activity. The dynamics of agent population $n_i$ under constraints $\sum_{i}
n_i=1$ are described by stochastic differential equations (SDEs) derived from the master equation with
one-step transition $i \rightarrow j$ rates proposed by Kirman 
\cite{Kirman1993QJE}:
\begin{equation}
\mu_{ij} = \sigma_{ij} + h_{ij} n_j N, \label{eq:nonext}
\end{equation}
where $\sigma_{ij}$ describes the individualistic switching tendency,
and $h$ quantifies influence of peers ($n_j N$). Note that a symmetric
relation $h_{ij} = h_{ji}$ is usually assumed and in the case of pairwise global coupling of agents number of peers is proportional to the total number of agents $N$. A basic understanding
of financial market dynamics allows us to make assumptions that simplify
the model.

We first assume that the three states correspond to three trading
strategies: fundamental ($f$), optimistic ($o$), and pessimistic
($p$), thus $i$ may take values $f$, $o$ and $p$. Fundamental traders assume that the price will approach a
fundamental price $P_f$ that is determined purely by market fundamentals.
Optimistic and pessimistic trading are two opposite approaches in the same chartist
($c$) trading strategy, i.e., optimists always buy and pessimists always
sell. Mathematical forms of the excess demands, $D_i$, for both
fundamental and chartist strategies are given by
\cite{Alfarano2005CompEco}
\begin{eqnarray}
& D_f = n_f \left[ \ln P_f - \ln P(t) \right] , \\ 
& D_c = r_0 (n_o - n_p) = r_0 n_c \xi , 
\end{eqnarray}
where $P(t)$ is the current market price of an asset, $r_0$ the relative
impact of chartists, and $\xi = \frac{n_o - n_p}{n_c}$ the average
mood. These three trading strategies are also considered in numerous other similar approaches \cite{Lux1999Nature,Alfarano2005CompEco,Samanidou2007RepProgPhys,Cincotti2008CompEco,Feng2012PNAS}.
Furthermore fundamentalist trading strategy, as described here, may be related to the concept of ``rational'' agents as used in
\cite{Lo2004JPM, Lo2005JIC,Galam2016Chaos, Dhesi2016Chaos}, while chartists, both optimists and pessimists,
are mostly equivalent to ``maladapted'' agents in \cite{Lo2004JPM, Lo2005JIC,Galam2016Chaos, Dhesi2016Chaos}.

 Combining $D_f$ and $D_c$, we obtain the log-price 
\cite{Alfarano2005CompEco,Gontis2014PlosOne},
\begin{equation}
p(t) = \ln \frac{P(t)}{P_f} = r_0 \frac{n_c}{n_f} \xi = r_0
\frac{1-n_f}{n_f} \xi . \label{eq:pdefin} 
\end{equation}

We next simplify the model by assuming that optimists and pessimists are
high-frequency trend followers, i.e., chartists. Chartists trade among
themselves $H$ times more frequently than with fundamentalists. There is
no genuine qualitative difference between optimists and pessimists in
terms of herding interactions, and certain symmetric relationships are
thus implied ($\sigma_{op}=\sigma_{po}=\sigma_{cc}$ and $h_{op}=H
h_{fc}=H h$). Chartists share their attitude towards fundamental trading
($\sigma_{pf}=\sigma_{of}=\sigma_{cf}$) and fundamentalists are
indifferent to arbitrary moods ($\sigma_{fp}=\sigma_{fo}=\sigma_{fc}/2$
and $h_{fp}=h_{fo}=h$). The assumption that fundamentalists are
long-term traders and chartists short-term traders can be written as ($H
\gg 1$, $\sigma_{cc} \gg \sigma_{cf}$ and $\sigma_{cc} \gg
\sigma_{fc}$). Under these assumptions the dynamics is well
approximated by two nearly independent SDEs
\cite{Kononovicius2013EPL,Gontis2014PlosOne} that resemble the original
SDE from the two-state herding model
\cite{Kirman1993QJE,Alfarano2005CompEco},
\begin{eqnarray}
& \rmd n_f = \frac{(1-n_f) \varepsilon_{cf} - n_f
    \varepsilon_{fc}}{\tau(n_f)} \rmd t + \sqrt{\frac{2 n_f
      (1-n_f)}{\tau(n_f)}} \rmd W_{f} , \label{eq:nftau}\\ 
& \rmd \xi = - \frac{2 H \varepsilon_{cc} \xi}{\tau(n_f)} \rmd t +
  \sqrt{\frac{2 H (1-\xi^2)}{\tau(n_f)}} \rmd W_{\xi} , \label{eq:xitau} 
\end{eqnarray}
where $\tau(n_f)$ is the inter-trade time, and $W_{f}$ and $W_{\xi}$ are
independent Wiener processes. Equations (\ref{eq:nftau}--\ref{eq:xitau}) can be derived starting from the 6 one step transition probabilities and corresponding master equation, see \cite{Kononovicius2013EPL} for details, or just using adiabatic approximation in the description of optimist-pessimist dynamics as in \cite{Gontis2014PlosOne}.  Note that in the above equations we scale
model parameters, $\varepsilon_{cf} = \sigma_{cf} / h$,
$\varepsilon_{fc} = \sigma_{fc} / h$, and $\varepsilon_{cc} =
\sigma_{cc} / (H h)$, as well as time $t_s = h t$ (omitting the
subscript $s$ in the equations).

We consider the inter-trade time $\tau(n_f)$ a macroscopic feedback
function, which can take the form
\begin{equation}
\frac{1}{\tau(n_f)}= \left( 1 + a_{\tau} \left| \frac{1-n_f}{n_f}
\right| \right)^{\alpha}. \label{eq:taunfxi} 
\end{equation}

This form is inspired by empirical analyses
\cite{Gabaix2003Nature,Farmer2004QF,Gabaix2006QJE,Rak2013APP}, where the
trading activity is proportional to the square of the absolute returns
(thus $\alpha=2$). This form depends on the long-term component of
returns in the proposed model (see \cite{Kononovicius2012PhysA}) and
$\frac{1}{\tau(n_f)}$ converges to unity when $n_f$ approaches 1. The
trading activity never reaches zero, and in non-volatile periods it
fluctuates around some equilibrium value. Note that in this approach $\tau(n_f)$ implements the macroscopic feedback based on the pairwise  global herding interaction of agents through their exchange in the pairwise trade action, see previous papers \cite{Kononovicius2012PhysA,Kononovicius2013EPL,Gontis2014PlosOne} for more details. 

Note that present form of Eq. (\ref{eq:taunfxi}) is slightly different from the previously published in \cite{Gontis2014PlosOne} as here we take off the dependence on high frequency fluctuations $\xi$ and parameter value $a_{\tau}$ will be slightly different from $a_0$. This simplification is very important as it makes Eq. (\ref{eq:nftau}) independent from Eq. (\ref{eq:xitau}) and provides much more transparent interpretation of the model and results. This minor change of the model conditions some change of the other parameter values.
 
Equations (\ref{eq:nftau}--\ref{eq:taunfxi}) constitute the complete
set for the macroscopic description of endogenous agent dynamics and
together with Eq.~(\ref{eq:pdefin}) constitute a model of financial
markets. Model simulation is based on numerical solution of Eqs. \eqref{eq:nftau} and \eqref{eq:xitau}. 

Distinctive feature of this particular approach is its analytical tractability in the form of SDEs (\eqref{eq:nftau}-\eqref{eq:xitau}). As was shown in \cite{Kononovicius2012PhysA}, Eq. (\eqref{eq:nftau}) written for the new variable $x$ in the region of high values of variable belongs to the class of nonlinear SDE's,  reproducing power-law statistics: PDF and PSD \cite{Kaulakys2005PhysRevE,Ruseckas2014JSM}. Furthermore, these equations exhibit a fascinating scaling property \cite{Ruseckas2014JSM}: the scaling of variable $x_s=ax$ is equivalent to the scaling of time $t_s=a^{2(\eta-1)}t$, where $\eta$ is the exponent of multiplicative noise term. This lies in the background of relation between power-law stationary PDF, $P(x) \sim x^{-\lambda}$, and PSD of $x$, $S \sim f^{\beta}$, where the general class of SDE, just with two parameters $\lambda$ and $\eta$ together with related exponent of PSD for $x$ , $\beta$, can be written as

\begin{equation}
\rmd x = (\eta-\frac{\lambda}{2})x^{2 \eta -1} \rmd t + x^{\eta} \rmd W, \quad \beta = 1+\frac{\lambda-3}{2(\eta-1)}. 
\label{eq:beta}
\end{equation}
 
The necessary condition for Eq. \eqref{eq:beta} is $\eta \neq 1$, see Eqs. (8, 9) in \cite{Ruseckas2010PhysRevE} for the corresponding Fokker-Planck equation and its steady-state solution. Models in finance usually consider the case $\eta < 1 $ and only rarely
the case $\eta > 1 $ \cite{Jeanblanc2009Springer}. Our herding based consideration belongs to the second one with the best fit to the empirical data in the region $3/2 \leq \eta \leq 5/2$. Note that introducing variable trading activity of agents into Eq. (\ref{eq:nftau}) \cite{Kononovicius2012PhysA}, we strengthen the non-linearity  of the basic stochastic differential equation (\ref{eq:beta}),
increasing the exponent of multiplicativity $\eta$.  The SDE (\ref{eq:beta}) exhibits nearly the same statistical properties as proposed endogenous model considered without high frequency fluctuations of the chartists $\xi(t)$. The main parameters of this power-law behavior can be written as follows \cite{Kononovicius2012PhysA}:

\begin{eqnarray}
\eta=\frac{3+\alpha}{2}, \quad
\lambda=\varepsilon_{cf}+\alpha+1, \quad
\beta=1+\frac{\varepsilon_{cf}+\alpha-2}{1+\alpha}.
\label{eq:parameters}
\end{eqnarray}

As in this simplified representation of the model $x$ has a meaning of the long-term absolute return (volatility), its power law behavior is very informative about statistical properties of the proposed model. For example, the contribution of  introduced feedback on trading activity may be recovered from the dependence of power-law exponents on $\alpha$, see Eqs. (\ref{eq:parameters}).

Understanding of the self-similarity and the long range dependence observed in the financial markets is usually based on the fractional Brownian motion \cite{Baillie1996JE,Giraitis2007,Giraitis2009}. Here we argue that the class of nonlinear stochastic differential equations (\ref{eq:beta}) can serve as an alternative mechanism explaining the property of the long range dependence in the financial markets. 

From our point of view, there are too many models based only on the endogenous dynamics of agents. First of all they are not realistic enough and in our approach it is impossible to adjust the both exponents of absolute return power-law behavior $\lambda$ and $\beta$ to the empirical data with the same set of parameters. For the more realistic model it is necessary to combine exogenous and endogenous fluctuations of the markets. As exogenous one we consider  the noise of 
 order flow fluctuations.

We substitute the endogenous price $p(t)$, Eq.~(\ref{eq:pdefin}),
calculated using Eqs.~(\ref{eq:nftau}--\ref{eq:taunfxi}) for $n_f$ and
$\xi$, into Eqs.~(\ref{eq:return}--\ref{eq:defvolatil}) to complete
the model, which now includes the endogenous and exogenous
fluctuations. It has been demonstrated \cite{Kononovicius2015PhysA} that the
model now resembles versions of non-linear GARCH(1,1) models
\cite{Engle1986ER,Higgins1992IER}. The advantage of agent-based models over pure stochastic models is that
their parameters are more closely related to real-world scenarios and
real human behavior. 

In the following we analyze the one minute, daily and monthly recorded time series. In
numerical simulations we set 1/390th of a trading day as the smallest
tick size $\delta$, and individual returns are calculated between these
ticks. We calculate the returns for long time periods $\Delta$, e.g., one day, by
summing up the consecutive short-time returns $r_{\delta}(t)$.

To account for the daily pattern observed in real data in NYSE and
FOREX, we introduce a time dependence \cite{Gontis2014PlosOne} into
parameter $b_0$, i.e., 
\begin{equation}
b_0(t)=b_0 \exp [-(\{t \mathrm{mod} 1 \}
  -0.5)^2/w^2]+0.5,
\label{eq:b0}
\end{equation}
where $w$ quantifies the width of intra-day
fluctuations. Although the model is designed to reproduce the power-law
behavior of absolute returns PDF and PSD, it also reproduces
the statistical features observed in
volatility return intervals.

\section{Results}
\begin{figure}[ht]
\centering
\includegraphics[width=.4\textwidth]{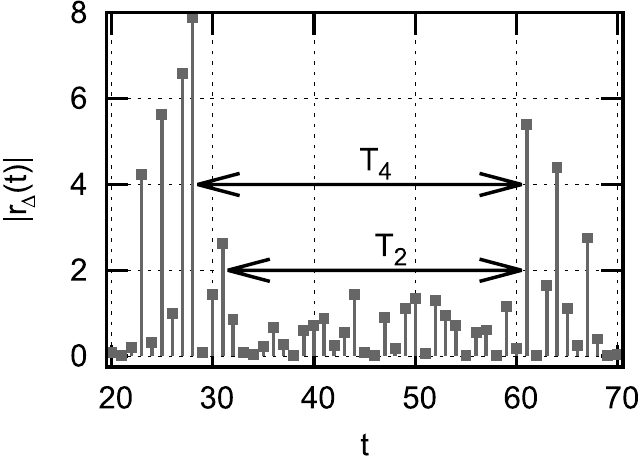}
\caption{The definition of return intervals $T_q$. Return intervals $T_q$ between the
  volatilities of the price changes that are above a certain threshold
  q, measured in units of standard deviations of returns (not absolute
  returns). Here two values of threshold $q=2$ and $q=4$ are shown in
  the time series of absolute return.}
  \label{fig:1}
\end{figure}

In this study we analyze the empirically established statistical
properties of volatility return intervals in financial markets
\cite{Yamasaki2005PNAS,Wang2006PhysRevE,Wang2008PhysRevE} and use the
same definition of this financial variable shown in
Fig.~\ref{fig:1}.

For two absolute return threshold values $q=2$ and $q=4$ the return
intervals are $T_2$ and $T_4$, respectively. They measure the time
intervals between consecutive spikes of absolute returns that exceed
threshold value $q$, measured in units of standard deviation of the returns in
the time series of the specific asset.

\subsection{PDF and PSD of absolute return}
We test how well the model reproduces the empirical PDF and PSD of
returns for NYSE stocks and FOREX exchange rates across a wide range of
time intervals $\Delta$ that range from 1/390th to 1 trading day. We set
the model parameters to be $\delta=1/390$ day $=3.69$ min., which is equivalent to 1 NYSE
trading minute, $\varepsilon_{cf}=1.1$ and $\varepsilon_{fc}=3$, which
define the anti-symmetric distribution of $n_f$, $\varepsilon_{cc}=3$,
which ensures the symmetric distribution of $\xi$, $H=1000$ which
adjusts the PSDs of the empirical and model time series, $a_0=1$ and
$a_{\tau}=0.7$, which are empirical parameters defining the sensitivity
of market returns and trading activity to the populations of agent
states, $\alpha=2$, which is selected based on the empirical analyses
\cite{Gabaix2003Nature,Farmer2004QF,Gabaix2006QJE,Rak2013APP} and our
numerical simulations confirm this choice as well, and $h=0.3\times10^{-8}
s^{-1}$, which is the main time-scale parameter that adjusts the model
to fit the real time-scale.  All the parameter values are kept constant
throughout the analysis that follows.

\begin{figure}[ht]
\centering
\includegraphics[width=.9\textwidth]{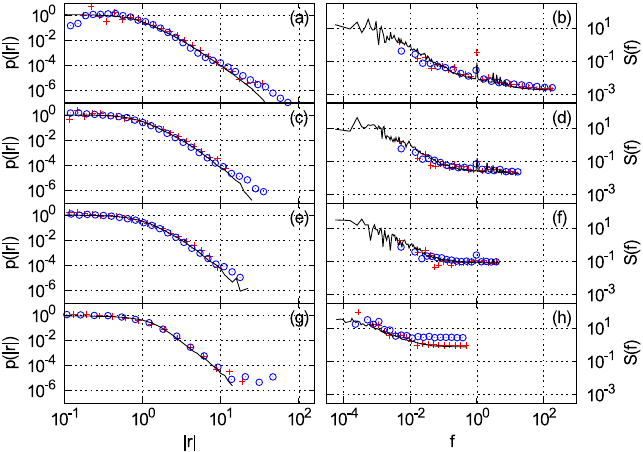}
\caption{A comparison between theoretical and empirical stationary PDFs and PSDs of absolute return. Theoretically calculated results - black lines, empirical results for NYSE stocks--circles and FOREX exchange rates--pluses. Stationary PDFs: (a), (c), (e), (g) and PSDs: (b), (d), (f), (h). (a) and
(b) for time-scales $\Delta=1/390$ trading day ; (c) and (d) -- $\Delta=1/39$ trading day ; (e) and (f)-- $\Delta=4/39$ trading day, frequencies in PSD graphics are given in 1/(1 trading day). Results for time scales $\Delta=$ one trading day of all considered assets from NYSE (circles) and of 10 exchange rates from FOREX (pluses) are in (g) and (h), where empirical series are from 1962 to 2014 year for NYSE series and from 1971 to 2014 year for FOREX. Model parameters are set as follows: $h=0.3 \times 10^{-8}s^{-1}$;
  $\delta = 3.69$ min.; $\varepsilon_{cf}=1.1$; $\varepsilon_{fc}=3$;
  $\varepsilon_{cc}=3$; $H=1000$; $a_0=1$; $a_{\tau}=0.7$;
  $\alpha=2$ for both NYSE and FOREX.}
  \label{fig:2}
\end{figure}

Figs.~\ref{fig:2}(a)--\ref{fig:2}(f) compare high
frequency NYSE and FOREX empirical data with the results of the model: numerical solution of Eqs. (\ref{eq:return},\ref{eq:defvolatil},\ref{eq:pdefin},\ref{eq:nftau},\ref{eq:xitau},\ref{eq:taunfxi},\ref{eq:b0}), see \cite{Gontis2014PlosOne} for details.
The data comprise a set of 26 stocks traded for 27 months from January
2005 and the USD/EUR exchange rate during a 10-year period beginning in
2000, and the empirical return series are normalized using return
standard deviation $\sigma_{\Delta}$.
Figs.~\ref{fig:2}(a)--\ref{fig:2}(f) show that the
model results are in a good agreement with the high frequency empirical
PDFs and PSDs.

\subsection{Contribution of various noises into the statistics of return intervals}
The heuristic model of volatility was designed to reproduced first and second order statistics of absolute return in the financial markets \cite{Gontis2014PlosOne}. The idea was to find the most simple version of consentaneous  agent based and stochastic model capable to reproduce PDF and PSD of absolute return observed for various financial markets and assets. It means that we normalize all empirical return data by standard deviation to the same PDF of absolute return first and then define the set of model parameters to reproduce empirical (stylized) PDF and PSD with all peculiarities. This procedure more relies on the understanding of statistical properties arising from the class of stochastic differential equations (\ref{eq:beta}) than on formal econometric procedures such as  GMM or SMM. Such stylized peculiarities as PSD with two different  values of exponent $\beta$ and spikes related to seasonality make the model much less appropriate for the formal consideration. The major achievement of such approach is ability to reproduce the same scaling of model and stylized statistical properties in very wide range of time windows $\Delta$.

Having such as simple as possible, but sophisticated enough model of absolute return, we demonstrate the capability of this model with the same set of parameters to reproduce a new class of empirical statistical properties: unconditional and conditional PDFs of high volatility return intervals. First of all, we demonstrate that all noises included into this model contribute to the PDF of absolute return intervals. As a first step, we analyze the long-term chartist fundamentalist dynamics, which can be described by  ratio $x=n_c/n_f=(1-n_f)/n_f$ defined by Eq. (\ref{eq:nftau}) and having statistical properties arising from Eq. (\ref{eq:beta}), which can be derived from Eq. (\ref{eq:nftau}) in the region of high $x$ values. Note that this is the main constituent of the long-term return fluctuations.  Second, we switch on exogenous noise, but keep $\xi$ and $b_0$ constant. This allows us to investigate the interaction of the long-term endogenous dynamics $x$ with exogenous noise by analyzing $\mid r_\delta(t) \mid$ and $\mid r_\Delta(t) \mid$. Third, we switch on optimist-pessimists dynamics $\xi(t)$ and analyze the absolute return series, keeping $b_0$ constant. And finally, we switch on intraday fluctuations and analyze full model with $b_0$ defined by Eq. (\ref{eq:b0}).

\begin{figure}[ht]
\centering
\includegraphics[width=.9\textwidth]{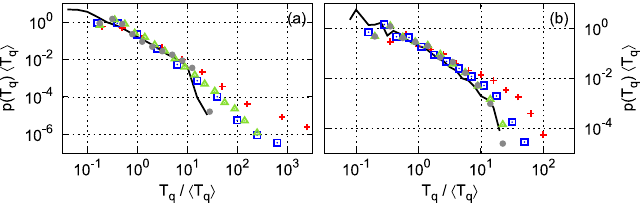}
\caption{Contribution of various noises into the PDF of absolute return intervals. (a) Scaled PDF of $T_q$ for the return definition time window $\Delta=\delta=1/390$ trading day; (b) scaled PDF of $T_q$ for the $\Delta=1$ trading day. Red pluses - only chartist fundamentalist dynamics $x$; blue squares -   fundamentalist dynamics $x$ with exogenous noise switched on; green triangles - fundamentalist and chartist joint dynamics
$x \xi$ with exogenous noise switched on; gray circles - full scale model with seasonality included; black line - the empirical PDF calculated from normalized series of NYSE stocks. All parameters of the model are the same as in previous figure, value of threshold $q=2.0$.}
\label{fig:3}
\end{figure}

Fig.~\ref{fig:3} compares the scaled PDFs of absolute return intervals $T_q$ calculated with four different compositions of the model and for empirical data of NYSE stocks.  In both sub-figures full model PDF of $T_q$ is in a good agreement with empirical data and one can observe considerable deviations from empirical data when part of noises is excluded from the model. In sub-figure (a), where $\Delta=\delta$, the contribution of optimist-pessimists dynamics $\xi(t)$, looks less noticeable as frequency of exogenous fluctuations is much higher than of $\xi(t)$ and of $x(t)$ fluctuations, nevertheless, the contribution of other noises is noticeable very well. In sub-figure (b), where $\Delta=1$ trading day, PDFs of  $T_q$ are different for all four compositions of the model. These and other numerical results confirm that all fluctuations accounted in the proposed model are required to reproduce statistics of empirical return intervals.    
    
\subsection{Return intervals of high frequency return series}
\begin{figure}[ht]
\centering
\includegraphics[width=.9\textwidth]{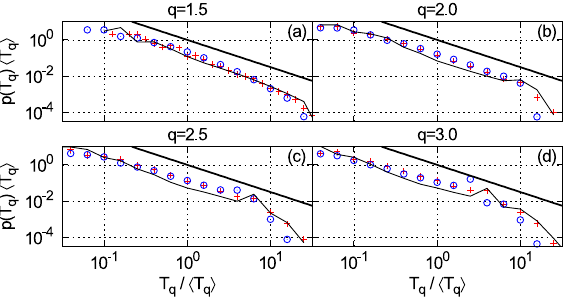}
\caption{A comparison between the model and empirical scaled unconditional PDFs of high frequency return intervals. Black lines - model PDFs; circles - the empirical PDFs calculated
\cite{Yamasaki2005PNAS,Wang2006PhysRevE} from normalized series of NYSE stocks and pluses - FOREX USD/EUR exchange rate. All parameters of the model are the same as in previous figure, values of thresholds $q$ are as follows: $1.5, 2.0, 2.5, 3.0$. The straight lines are shown to guide the eye showing a power-law with exponent $3/2$.}
\label{fig:4}
\end{figure}

\begin{figure}[ht]
\centering
\includegraphics[width=.9\textwidth]{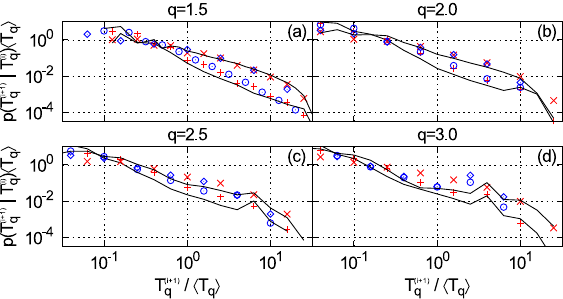}
\caption{A comparison between the model and empirical scaled conditional PDFs of high frequency return intervals. Black lines - scaled conditional PDFs of return intervals, $P(T_q^{i+1}\mid T_q^{i})$, circles and diamonds - the empirical PDFs calculated from normalized series of NYSE stocks, pluses and crosses - the empirical PDFs of USD/EUR exchange rate. Conditional PDFs are calculated with the same algorithm as in ref.~\cite{Wang2006PhysRevE,Yamasaki2005PNAS}, $T_q^{i} \leq Q_1$ - lower PDFs and $T_q^{i} \geq Q_8$ -  upper PDFs, where $Q_1$ and $Q_8$ are $1/8$ and $7/8$ quantiles of $T$ sequence accordingly. All parameters of the model are the same as in previous figures, values of thresholds $q$: $1.5, 2.0, 2.5, 3.0$.}
  \label{fig:5}
\end{figure}

Our goal now is to explain, using model, the statistical properties of the return
intervals of both stocks and currencies \cite{Yamasaki2005PNAS,Wang2006PhysRevE}.
Fig.~\ref{fig:4} compares the unconditional PDFs of the
model with the PDF obtained for 1/390th trading day returns of NYSE
stocks and USD/EUR exchange on FOREX, and Fig.~\ref{fig:5}
compares the conditional distribution functions. These results support the model showing that it 
successfully reproduces both unconditional and
conditional distribution functions.  When $q$ values are comparable with the returns from their power-law
part of PDF, $q>1.5$, the power-law behavior of return intervals prevails $P(T_q)\sim T_q^{-3/2}$. Notice that scaled unconditional PDFs of $T_q$ empirical as well as model given in Fig.~\ref{fig:4} are nearly the same for each value of $q$.
We do observe this power-law behavior with exponent $3/2$ in the model even when we simplify it by replacing the whole model by stochastic dynamics of $x=\frac{1-n_f}{n_f}$
defined in Eq.~(\ref{eq:nftau}) and the other noises are
switched off. The cutoff of this power-law behavior for high values of $T_q$ appears when other noises are switched on again. Our numerical simulations of the model show that for values of $q$, comparable with returns in very tail of their power-law PDF, the exogenous
noise in Eq.~(\ref{eq:return}) is responsible for the deviations from
$3/2$ law, when $\xi$ as well as intra-day trading activity dynamics
force the scaled PDF back to a power-law $3/2$ behavior.  Such impact of the exogenous
noise increases with higher values of time window $\Delta$.

For the threshold value $q=1.5$, when $T_q^{i} \leq Q_1$ and $T_q^{i}
\geq Q_8$ the conditional distribution functions $P(T_q^{i+1}\mid
T_q^{i})$ are clearly different, indicating that there is a memory
effect. Here $i$ is the index in the consecutive $T_q$ sequence, $Q_1$ the 1/8th quantile and
$Q_8$ the 7/8th quantile of $T_q$ series. When threshold values are higher the
conditional PDFs become closer and might overlap.  Our numerical modelings confirm that the necessary condition
for this memory effect is the presence of long term dynamics,
Eq.~(\ref{eq:nftau}), and exogenous noise, Eq.~(\ref{eq:return}). The
speculative dynamics $\xi$ and intra-day seasonality contribute to the
dynamic behavior of the system, the persistence of a 3/2 power-law, and
the memory effects. Note that all noises defined by the model are
reflected in the PDFs of the volatility return intervals. The intraday fluctuations accounted in the model by Eq. (\ref{eq:b0}) contribute to the high frequency conditional PDFs of return intervals, see Fig. \ref{fig:5}, and help to achieve qualitative agreement with empirical data. Nevertheless, we have to acknowledge that the method we use to account the intraday fluctuations is oversimplified and some quantitative deviations from empirical data are present for the higher threshold $q$ values.

Our results support the empirical finding  \cite{Yamasaki2005PNAS,Wang2006PhysRevE,Wang2008PhysRevE} that the
PDF of the return intervals can be scaled to the same form common for
different thresholds $q$.  Note that the difference in scaling exponent
between what we obtained ($3/2$) and that obtained (2) in previous
research is related to the use of different procedures for the return
normalization, which in turn leads to different threshold choices.  The
thresholds used in previous papers are considerably lower than the ones
we use in our model simulations and are outside the power-law portion of the
return PDF.  Because the contribution of the main SDE in
Eq.~(\ref{eq:nftau}) prevails over other noises only in the power-law
portion of the return PDF, we choose higher values for threshold $q$ and
also show the deviation from the $3/2$ law for $q=1.5$. This lowest
value, $q=1.5$, demonstrates the transition to the regime in which the
return intervals are extremely short and the dynamic complexity of the
signal extremely high. We cannot consider the high frequency fluctuations
in this regime as caused by a one-dimensional stochastic process because
other noises are also contributing.  Thus the exponent of the return
intervals tends to values higher than $3/2$. The empirical studies of
return interval statistics described in
Refs.~\cite{Bogachev2007PRL,Bunde2011EPL,Bunde2014PRE} demonstrate the
transition from a $3/2$ power-law to the exponential distribution of the
unconditional PDF. Note that the authors of these studies also select
lower values for the thresholds.

\subsection{Return intervals of daily return series}
We next analyze the daily returns data of 10 NYSE stocks obtained from Yahoo Finance, and also the USD
historical exchange rates with currencies AU, NZ, POUND, CD, KRONER,
YEN, KRONOR, and FRANCS traded on FOREX and obtained from the Federal
Reserve. We first determine the appropriate scaling of the daily series
of returns in the FOREX and NYSE exchanges. Because it is unlikely that
those return series that exceed 50 years will be stationary, we
normalize them by using a moving standard deviation procedure with a 5000-day
time window. Each time series of all assets in both markets is
normalized using this procedure. Fig.~\ref{fig:2}(g) compares
the normalized empirical PDFs with the model PDF, and
Fig.~\ref{fig:2}(h) shows the PSDs. Notice that PSD of stock absolute returns in high frequency area has a slightly higher value than model and currency exchange PSDs. There is good agreement of PSDs in low frequency area.

\begin{figure}[ht]
\centering
\includegraphics[width=.9\textwidth]{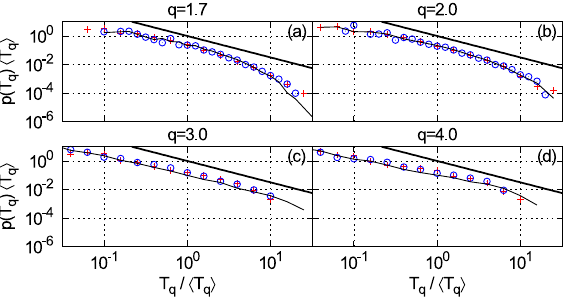}
\caption{A comparison between the model and empirical scaled unconditional PDFs of daily return intervals. Black lines - model unconditional PDFs of return intervals; circles -  empirical PDFs calculated from normalized return series of NYSE stocks; pluses - empirical PDFs calculated from normalized return series of currency exchanges. All parameters of the model are the same as in previous figures, values of thresholds are as follows: $1.7, 2.0, 3.0, 4.0$. The straight lines are shown to guide the eye showing a power-law with exponent $3/2$.}
  \label{fig:6}
\end{figure}

Fig. \ref{fig:6} shows that the unconditional PDFs of the
daily scaled return intervals for NYSE stocks and FOREX exchange rates coincide for each threshold value.  Note
that in both NYSE and FOREX markets the unconditional PDFs agree with the
model PDFs. This indicates a high degree of scaling in the return
intervals. The theoretical framework provided by our model is able to
explain this scaling.  Note that for the highest threshold value $q=4$
the power-law exponent of the unconditional PDF in
Fig.~\ref{fig:6} deviates from $3/2$ and approaches $1$.

\begin{figure}[ht]
\centering
\includegraphics[width=.9\textwidth]{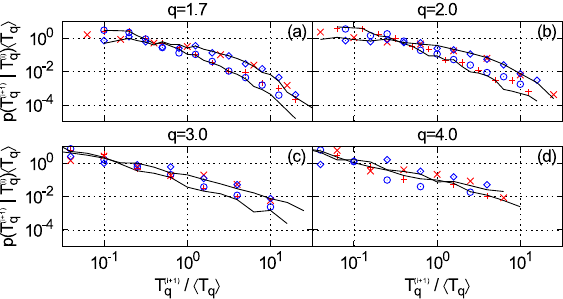}
\caption{A comparison between the model and empirical scaled conditional PDFs of daily return intervals. Black lines - model conditional PDFs of daily return intervals, $P(T_q^{i+1}\mid T_q^{i})$ with $T_q^{i} \leq Q_1$ and $T_q^{i} \geq Q_8$ groups; circles and diamonds - the corresponding empirical PDFs calculated from normalized series of NYSE stocks; pluses and crosses - the corresponding empirical PDFs calculated from normalized series of currency exchanges. All parameters of the model are the same as in previous figures, values of thresholds are as follows: $1.7, 2.0, 3.0, 4.0$.}
\label{fig:7}
\end{figure}

Fig.~\ref{fig:7} shows that the conditional PDFs of the
model agree with the conditional PDFs of the daily volatility return
intervals records of both the NYSE and FOREX markets. When we increase the
threshold, the conditional PDFs become closer and seem to overlap in both the empirical data and
the model results, but we can not rule out that the seemingly overlap is due to the increased level of noise for high q.

\subsection{Return intervals of monthly series for S\&P500 index} 
\begin{figure}[ht]
\centering
\includegraphics[width=.9\textwidth]{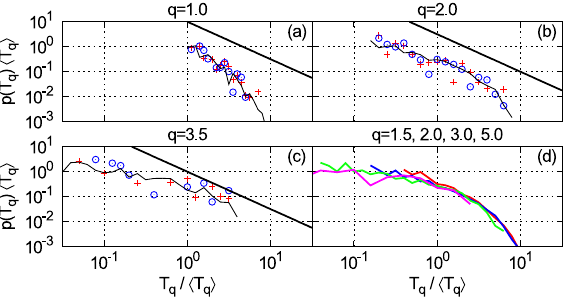}
\caption{A comparison between the model and empirical scaled unconditional PDFs of monthly $S\&P500$ return intervals. Black lines - model PDFs of monthly return intervals; circles - the empirical PDFs calculated from normalized series of historical $S\&P500$ data; pluses - the inflation adjusted $S\&P500$ series represented as real price. The straight lines are shown to guide the eye showing a power-law with exponent $3/2$. All parameters of the model are the same as in previous figures, values of thresholds $q$ are as follows: (a) - 1, (b) - 2, (c) - 3.5. In subfigure (d) - numerical calculations of unconditional scalded PDFs for four values of threshold $q$: 1.5 (red), 2 (blue), 3
  (green), 5 (purple).}
  \label{fig:8}
\end{figure}

We use data from an S\&P500 monthly series spanning a 145-year period
provided by Shiller \cite{Shiller2015IE} to demonstrate the behavior of
return intervals for extremely long time-scales.
Fig.~\ref{fig:8} shows that the above model, which reproduced the statistics of high frequency data, successfully mimics the
PDFs of the volatility return intervals for even the longest time
scales. We plot the empirical PDFs of return intervals for a nominal
S\&P500 and inflation adjusted series and compare them with the 
model series. The chosen threshold values range from 1.0 to 3.5 and
represent several exponents of PDF.  In the longest time-scales, the
return interval distribution deviates from the $3/2$ power law for both
the lowest and highest threshold values.  Although the number of data
points is limited, the model is able to capture these deviations and
reproduce the behavior of index data.

\subsection{Deviations from the 3/2 law}
Because our model reproduces the statistical properties of empirical
data for a wide range of assets and time-scales, we can use it to
explain why increasing threshold $q$ causes deviations from the
theoretical 3/2 power law and the seemingly absence of memory in the conditional
PDFs of return intervals.  In particular, the model allows us to
gradually switch off various noises and analyze how this changes the
statistical properties of the return intervals.
 
The model conditional PDFs and the empirical data conditional PDFs
overlap at approximately the same threshold values at which the
unconditional PDFs deviate from the $3/2$ power law, for example, see Fig.~\ref{fig:6}(d) and Fig.~\ref{fig:7}(d). This phenomenon is
stronger for larger time $\Delta$ scales, and we see no memory effects
in the empirical S\&P500 monthly series.  Fig.~\ref{fig:8}(d)
shows the unconditional PDFs calculated numerically for several values
of $q$, which resemble the exponential function discussed in
Ref.~\cite{Yamasaki2005PNAS} and obtained by reshuffling the absolute
return time series. This indirectly confirms that the volatility return
intervals for the S\&P500 historical time series display no memory effect.

Our numerical simulations of the model suggest that the primary cause of the $3/2$
power-law behavior of the return intervals is the long-term SDE (see
Eq.~(\ref{eq:nftau})). Other dynamic processes such as the speculative
mood $\xi$ (see Eq.~(\ref{eq:xitau})) and the intra-day seasonality
contribute to the stability of this phenomenon. Although the exogenous
noise in Eq.~(\ref{eq:return}) causes the unconditional PDFs to deviate
from the 3/2 power law, this noise is a necessary condition for the
memory effect to emerge in conditional PDFs. From our numerical
simulations we conclude that the deviations from the $3/2$ power-law and
disappearance of the memory effect occur when the stochastic component
is stronger than the dynamic component. This process of domination
occurs when the threshold value is so high that the dynamic processes
cannot reach it when the noise is switched off. Note that threshold $q$
is measured in standard deviations of return, which grow approximately
as $\Delta^{1/2}$. Dynamic processes quantified in $\sigma_t$ of
Eqs.~(\ref{eq:pdefin}) and (\ref{eq:defvolatil}) with this set of model
parameters can only approach the threshold when its value is
approximately equal to the standard deviation of the daily return time
series. Thus when the thresholds are much higher the dynamic component
is weaker than the stochastic component and the return intervals begin
to deviate from the $3/2$ power law. The prevailing stochastic nature of
the return time series destroys the memory effect, which requires that
both dynamic and stochastic components be in the system.

\section{Discussion}

We have observed scaling and memory properties in the volatility return
intervals in empirical data from the NYSE and the FOREX
\cite{Yamasaki2005PNAS,Wang2006PhysRevE,Wang2008PhysRevE}.  Our model is in agreement with the empirical return intervals that scale with the mean return interval
$<T>$ as $P_q(T)=<T>^{-1}f(T/<T>)$. The scaling function $f(x)$ is
consistent with the power-law form $f(x)\sim x^{-3/2}$, which arises
from the general theory of first-passage times in one-dimensional
stochastic processes \cite{Redner2001Cambridge,Jeanblanc2009Springer}. We recover the same
scaling form for all assets analyzed from the NYSE and FOREX markets for
return definition times $\Delta$ ranging from one minute to one month
and for a wide range of thresholds $q$, which represent the power-law
component of the empirical return series. Our model also captures the
deviations of the volatility return interval PDF exponent from the main
value $3/2$ and explains the origin of these deviations.
 
We also have observed that at low $q$ values at the beginning of
the power-law component of the empirical return series, for both
one-minute and one-day periods, the conditional PDFs $P(T_q^{i+1}\mid
T_q^{i})$ for $T_q^{i} \leq Q_1$ and $T_q^{i} \geq Q_8$ are different,
and this indicates the presence of a memory effect. Our model suggests
that this effect is caused by a complex interplay of all the noises
included in the system.  The necessary condition for the memory effect
is the presence of long-term agent dynamics and exogenous noise. High
threshold $q$ values seem to cause the memory effect to disappear as the
stochastic component of the volatility begins to prevail.

When we compare the results of our model with the monthly data from the
S\&P500 we are able to extend our research on the scaling properties of
the volatility return interval up to the natural limits of the
phenomenon. We thus suggest that the deviations of the PDF exponent
from $3/2$ are caused by an interplay between agent dynamics and
exogenous noise. The standard deviations of return in the S\&P500
monthly series are so high that the dynamic component of the system
becomes negligible and the stochastic component dominates. This causes
the exponential scaling functions of the return intervals and the
disappearance of the memory effects.

We have found that the statistical and scaling properties such as the
observable power-law behavior in the returns can be explained using non-linear stochastic
modeling \cite{Gontis2014PlosOne}. The extreme power-law scaling
properties observed in all assets, markets, and time-scales can be explained by the
scaling properties of a class of nonlinear stochastic
differential equations described in detail in
Refs.~\cite{Kaulakys2005PhysRevE,Ruseckas2014JSM}. Our model here is
based on the herding interactions of agents, and its macroscopic version
is derived as a system of stochastic equations. These equations might be the origin for the power-law properties of the power spectral density
and signal autocorrelation represented by long-range
memory.
 
We have also demonstrated that the model can be scaled for markets with
trading hours of different durations and that the duration can be
extended to a 24-hour day. This allows a general approach to empirical
data scaling and the retrieval of the same power law properties in
different markets and different assets.

\section*{Acknowledgments}
This work was partially supported by Baltic-American Freedom Foundation
and CIEE. The Boston University work was supported by NSF Grants PHY 1444389, PHY 1505000, CMMI 1125290, and CHE-1213217,
and by DTRA Grant HDTRA1-14-1-0017 and DOE Contract DE-AC07-05Id14517.

%\bibliographystyle{elsarticle-num}
%\bibliography{return-intervals}

\end{document}